\documentclass[pre,10pt,amsmath,amssymb]{revtex4}
\usepackage{graphicx}

\usepackage{amsmath,amsfonts,amssymb,bm}

\begin{document}

\draft
\title{L\'evy flights and nonhomogenous memory effects: relaxation to a stationary state}

\author
{Tomasz Srokowski}

\affiliation{
 Institute of Nuclear Physics, Polish Academy of Sciences, PL -- 31-342
Krak\'ow,
Poland }

\date{\today}

\begin{abstract}
The non-Markovian stochastic dynamics involving L\'evy flights and a potential in the form of a harmonic 
and non-linear oscillator is discussed. The subordination technique is applied and the memory effects, 
which are nonhomogeneous, are taken into account by a position-dependent subordinator. In the non-linear 
case, the asymptotic stationary states are found. The relaxation pattern to the stationary state is derived 
for the quadratic potential: the density decays like a linear combination of the Mittag-Leffler functions. 
It is demonstrated that in the latter case the density distribution satisfies 
a fractional Fokker-Planck equation. The densities for the non-linear oscillator reveal a complex picture, 
qualitatively dependent on the potential strength, and the relaxation pattern is exponential at large time. 
\end{abstract} 


\maketitle


\section{Introduction}

The presence of long distribution tails in stochastic systems means a qualitatively different picture 
than a familiar Gaussian process. In particular, the variance of the process value $x$ 
diverges and the ordinary central limit theorem does not apply. 
Instead of the normal distribution, the power-law form of the asymptotic density distribution, 
$|x|^{-\alpha-1}$ ($0<\alpha<2$), emerges. Processes characterised by such tails are stable and called 
L\'evy flights. They are typical for complex systems and frequently observed in many areas 
of science \cite{shles,barn}. Processes involving L\'evy flights are well suited to describe phenomena 
far from equilibrium. Dynamics of a particle 
subjected to the L\'evy stable noise can be described by a Langevin equation, 
\begin{equation}
\label{la0}
dx(t)=-\frac{\partial}{\partial x}V(x)dt+\eta(dt),
\end{equation}
where $V(x)$ is a potential and the increments $\eta(dt)$ obey the L\'evy stable statistics 
with the stability index $\alpha$. Solution of the corresponding Fokker-Planck equation  
is simple for the linear case -- when $V(x)=$const, $\propto x$ 
or $\propto x^2$ -- and the resulting density reveals the same statistics as the driving noise \cite{jes}. 
That equation, in the presence of the L\'evy flights determined by a stability index $\alpha$, 
contains a fractional derivative defined by a Fourier transform \cite{old}, 
\begin{equation}
\label{frder}
\frac{\partial^\alpha}{\partial|x|^\alpha}f(x)={\cal F}^{-1}[-|k|^\alpha{\widetilde f}(k)]. 
\end{equation}
For the quadratic potential, the system converges with time to a stationary state. The stationary 
solutions in the asymptotic limit have been also found for strongly nonlinear systems, 
$V(x)\propto x^\gamma$ ($\gamma>2$), 
and the steep slope of the potential modifies the distribution making the variance finite \cite{che,che1}. 
This analysis was generalised to a two-dimensional problem \cite{szcz} and to systems driven by 
a multiplicative noise \cite{sro10}; the spectral analysis was performed for transport 
in an inhomogeneous medium \cite{kaz}. 

The above approach is Markovian (Eq.(\ref{la0}) is local in time) 
and does not take into account memory effects emerging in complex \cite{tak} 
and disordered systems \cite{bou} where, as a result of traps and faults, the particle 
experiences periods of long rests. Then the process can be formalised by a jump-size and waiting-time 
distribution in the framework of the continuous time random walk (CTRW) \cite{met}. The problem without 
any potential is well known: the variance -- existing for the Gaussian case -- 
rises slower than linearly, which means a subdiffusion, and the characteristic function of the density 
distribution decays according to the Mittag-Leffler pattern \cite{hau}. The diffusive behaviour for which 
the variance rises with time slower or faster than linearly is often called 
an anomalous transport, in contrast to a normal diffusion. Processes characterised by the anomalous transport 
-- e.g. the flow in porous media where fractional space derivatives model large motions
through highly conductive layers or fractures while fractional time derivatives 
describe motionless particles \cite{meer} -- 
can be conveniently handled by the subordination procedure \cite{sok,bar1,pir}. In that description, 
the Markovian dynamics proceeds in an operational time and is given by Eq.(\ref{la0}) (a subordinated 
process), whereas the memory effects are modelled by another Langevin equation 
for the physical time which is a non-decreasing random variable, determined by 
a one-sided density distribution $\xi$ (a subordinator). In the presence of 
a potential, the process (\ref{la0}) has a stationary limit and the subordinator determines a relaxation 
to that stationary state. Long tails of the waiting-time distribution mean a deviation from the exponential 
Debye relaxation. Non-exponential forms of the relaxation are common in complex systems: viscoelastic materials, 
synthetic polymers and biopolymers \cite{glo}. They emerge in supercooled liquids near the glass transition 
temperature and in amorphous polymers \cite{rich}; a typical decay form of the relaxation processes
is a Mittag-Leffler function (Cole-Cole) \cite{hil}. Diversity of systems exhibiting the non-exponential 
relaxation traces back to the universality present in complex systems: the empirical relaxation laws 
reflect a behaviour which is independent of the details of examined systems \cite{jon}. From that point of view, 
the non-exponential relaxation can be understood as a result of a strong interaction between diverse 
elementary units. 

In a nonhomogeneous medium, one can expect that the waiting time depends on the position. 
The corresponding CTRW description is coupled \cite{met}; its Markovian version contains the waiting-time 
distribution in a Poissonian form, characterised by a variable jumping rate \cite{kam}, and resolves itself 
to a Fokker-Planck equation with a multiplicative noise \cite{sro06}. 
In the disordered systems, many forms of the disorder may emerge \cite{bou} and 
the dynamics becomes complicated if one assumes that the trap positions slowly evolve with time 
(a quenched disorder), then the successive trapping times are correlated. 
The position-dependence of memory effects can be taken into account by applying  
a subordination technique and such a model has been recently proposed \cite{sro14}. 
One can expect that a large density of traps or regular structures in the phase space 
the trajectories stick to invoke a long resting time at specific regions. 
The subordination approach \cite{sro14} takes into account that position-dependence by 
introducing a variable intensity of the random generator of the physical time, given by 
a non-negative function $g(x)$. Then Eq.(\ref{la0}) describes the subordinated process 
and the dynamics is defined by the following set of equations, 
\begin{eqnarray}
\label{la}
dx(\tau)&=&F_d(x) d\tau+\eta(d\tau)\nonumber\\
dt(\tau)&=&g(x)\xi(d\tau), 
\end{eqnarray}
where the increments $\eta(dt)$ and $\xi(dt)$ obey the L\'evy stable statistics with the stability 
indexes $0<\alpha<2$ and $0<\beta<1$, respectively. The process $t(\tau)$ is non-decreasing
and biased by the medium properties: if, in particular, $g(x)$ is large, the particle stays at $x$ 
for a relatively long time, compared to the region where $g(x)$ is small. 
Eq.(\ref{la}) can be approximated by an ordinary subordination process if we 
decouple the random and the position-dependent ingredient; this can be accomplished 
by a two-step procedure \cite{sro14}. First, let us assume for the moment that $\xi$ is given by  
a one-sided density with finite moments (e.g. an exponential), approximate 
$\xi(d\tau)$ by the average, $\langle\xi\rangle d\tau$, and evaluate the operational time increment, 
$\Delta\tau=(\langle\xi\rangle g(x))^{-1}\Delta t$. Since the first equation (\ref{la}) can be discretized 
as $\Delta x=F_d(x) \Delta\tau+\Delta\eta\Delta\tau^{1/\alpha}$, the system (\ref{la}), after inserting 
$\Delta\tau$, resolves itself to the Langevin equation with a multiplicative noise 
in the It\^o interpretation which defines a Markovian process. The above equation incorporates 
the spatial distribution of traps but neglects the memory effects which are important if $\beta<1$. 
Then, as a second step, we take 
into account those effects subordinating the Markovian process to the random time determined by 
$\beta<1$. The final set of equations reads, 
\begin{eqnarray}
\label{las}
dx(\tau)&=&F_d(x)\nu(x) d\tau+\nu(x)^{1/\alpha}\eta(d\tau)\nonumber\\
dt(\tau)&=&\xi(d\tau), 
\end{eqnarray}
where $\nu(x)=[\langle\xi\rangle g(x)]^{-1}$; we assume in the following $\langle\xi\rangle=1$. 
The process described by Eq.(\ref{las}) was studied for $F_d=0$ \cite{sro14}: it corresponds to a Markovian CTRW 
with a variable jumping rate, subordinated to a random time, and resolves itself to a fractional 
Fokker-Planck equation with a variable diffusion coefficient. The system (\ref{las}) 
is used, for example, to describe such transport phenomena as advection, 
molecular diffusion and hydrodynamic dispersion, where the first equation (\ref{las}) 
is related to the advection-dispersion equation whereas the subordination accounts for memory \cite{berk12}. 

In the presence of an external potential, Eq.(\ref{la0}) predicts a convergence to a stationary density distribution. 
However, the stationary state is difficult to derive and known in its exact form only for few potentials \cite{che}. 
All the more difficult is a problem of the relaxation to that state and the time-dependent solutions are unknown 
even for the Markovian case. It is unclear, in particular, whether the Mittag-Leffler pattern is still valid 
when a process with a nonlinear deterministic force is subordinated to the random time. 
In the present paper, we address that relaxation problem and consider a more general case 
of the position-dependent memory. We find expressions for the asymptotic stationary density for the potential 
in a power-law form and derive the density for the subordinated process, as a function of the operational time, 
for a quadratic potential (Sec.II). Sec.III is devoted to the time-dependent solutions of Eq.(\ref{la}), (\ref{las}) 
where the relaxation function is derived and the influence of the nonhomogeneity effects 
on the decay rate of the density is discussed. 
 
\section{Stationary states and the subordinated process}

We consider a stochastic system under the influence of an non-linear deterministic force, 
\begin{equation}
\label{dfor}
F_d=-\lambda|x|^{\bar\gamma}\hbox{sign}(x),
\end{equation}
where $\lambda\ge0$ and $\bar\gamma=$const. This form of the stimulation is a natural generalisation of an ordinary 
harmonic oscillator and often discussed in connection with the chaotic dynamics \cite{che2}. 
Hence the first equation (\ref{las}) takes the form,  
\begin{equation}
\label{las1}
dx(\tau)=-\lambda|x|^{\bar\gamma}\nu(x)\hbox{sign}(x)d\tau+\nu(x)^{1/\alpha}\eta(d\tau), 
\end{equation}
and corresponds to a Fokker-Planck equation \cite{sch}, 
\begin{equation}
\label{frace}
\frac{\partial p_0(x,\tau)}{\partial \tau}=\lambda\frac{\partial}{\partial x}
[|x|^{\bar\gamma}\nu(x)\hbox{sign}(x)p_0(x,\tau)]+
\frac{\partial^\alpha[\nu(x) p_0(x,\tau)]}{\partial|x|^\alpha}, 
\end{equation}
where the fractional derivative is defined by Eq.(\ref{frder}). The equation 
for the characteristic function reads 
\begin{equation}
\label{fracek}
\frac{\partial}{\partial \tau}{\widetilde p_0}(k,\tau)=-\lambda k\frac{\partial}{\partial k}
{\cal F}[|x|^{\bar\gamma-1}\nu(x)p_0(x,\tau)]-|k|^\alpha{\cal F}[\nu(x)p_0(x,\tau)].
\end{equation} 

Eq.(\ref{fracek}) cannot be analytically solved, in general, and presence of the multiplicative 
factor $\nu(x)$ in the last term poses the main difficulty. We restrict our considerations 
to the asymptotic regime in positions, which corresponds to small wave numbers, and derive a stationary 
solution. Due to the factor $|k|^\alpha$ in the last term of Eq.(\ref{fracek}), the leading term 
in the solution is of the order $|k|^\alpha$ implying the asymptotic solution in a scaling form, 
\begin{equation}
\label{sol0nl}
p_0(x,\tau)=a(\tau)p_0(a(\tau)x)=a(\tau)^{-\bar\alpha}|x|^{-\bar\alpha-1}, 
\end{equation}
where $\bar\alpha=$const; the above solution is valid if $0<\bar\alpha<2$. 
Moreover, it is sufficient to keep only the lowest term, $k^0$, in the $k-$expansion 
of the Fourier transform argument and the transform resolves itself to the integral 
\begin{equation}
\label{ost}
\int\nu(x)p_0(x,\tau)dx. 
\end{equation}
For the evaluation of the argument of the first Fourier transform an additional assumption is necessary: 
we require that the asymptotic form of $g(x)$ is algebraic, $g(x)=|x|^\theta$, and that argument reads 
$a(\tau)^{-\bar\alpha}|x|^{\bar\gamma-\theta-\bar\alpha-2}$ for large $|x|$. 
The last term in Eq.(\ref{fracek}), in turn, follows from the integral (\ref{ost}), 
\begin{equation}
\label{ftr2}
-|k|^\alpha{\cal F}[\nu(x)p_0(x,\tau)]=-|k|^\alpha\langle\nu(x)\rangle(\tau), 
\end{equation}
and this expression is valid for any time. Now, we take the limit of large time and look for 
the stationary solution: $p_0(x,\tau\to\infty)=p_0^{(s)}(x)$ and $a(\tau\to\infty)=a_s$. 
In that limit, the first transform on the rhs of Eq.(\ref{fracek}) reads 
\begin{equation}
\label{ftr1}
f(k)=N^{-1}{\cal F}[Na_s^{-\bar\alpha}|x|^{-\delta-1}], 
\end{equation}
where $N=\langle|x|^{\bar\gamma-1}\nu(x)\rangle^{-1}$ and $\delta=-\bar\gamma+\theta+\bar\alpha+1$ 
($0<\delta<2$). 
The transform argument is normalised to unity and the transform itself can be simply evaluated up to 
the first order in $k$ by means of the following lemma \cite{gor}. If $W(x)$ is a normalised 
and symmetric density distribution and $\int_x^\infty dW(x')\sim b\alpha^{-1}x^{-\alpha}$ for 
$x\to\infty$ ($b>0$), then $1-\widetilde W(k)\sim bc|k|^\alpha$ for $k\to0$, where 
\begin{equation}
\label{gor}
c=\frac{\pi}{\Gamma(\alpha+1)\sin(\alpha\pi/2)}.
\end{equation}
Application of the above lemma to the present problem yields $f(k)=N^{-1}-ca_s^{-\bar\alpha}|k|^{\delta}$ and 
inserting of the evaluated expressions to Eq.(\ref{fracek}) produces Fourier transform from 
the stationary density which contains terms corresponding to $|k|^\alpha$ and $|k|^\delta$. 
The equation is satisfied for $\delta=\alpha$ and the parameter $\bar\alpha$ 
is determined: $\bar\alpha=\alpha+\gamma-1$, where $\gamma=\bar\gamma-\theta$. Moreover, we obtain 
\begin{equation}
\label{ast}
a_s=\left[\frac{\lambda\pi}{\langle\nu(x)\rangle\Gamma(\alpha)\sin(\alpha\pi/2)}\right]^{1/\bar\alpha}
\end{equation}
and the final asymptotic solution follows from Eq.(\ref{sol0nl}), 
\begin{equation}
\label{sol0nlf}
p_0^{(s)}(x)=\frac{\langle\nu(x)\rangle}{\lambda\pi}\Gamma(\alpha)\sin(\alpha\pi/2)|x|^{-\alpha-\gamma}. 
\end{equation}
The above solution exists if $\bar\gamma$ satisfies some conditions: it is normalisable for 
$\bar\gamma>1-\alpha+\theta$ and $\langle\nu(x)\rangle$ is finite for $\bar\gamma>1-\alpha$. 
The $x-$dependence of $p_0^{(s)}(x)$ is exact for any $\theta$ but $\langle\nu(x)\rangle$ 
depends on the density at small $|x|$ and cannot be determined without additional assumptions. 
For $\theta=0$, Eq.(\ref{sol0nlf}) is exact in a wide range of $\bar\gamma$, including negative values, 
and the condition $\bar\gamma>1-\alpha$ is required to get a stationary solution \cite{dyb}. 

Our aim is to establish how the system under the force field (\ref{dfor}) approaches the stationary 
state (\ref{sol0nlf}), i.e. we have to derive the time-dependence of the density $p_0(x,\tau)$. 
Its asymptotic form can be analytically derived for a linear case. 
From now on, we restrict our analysis to a power-law form of $g(x)$, 
\begin{equation}
\label{godx}
g(x)=|x|^\theta~~~~(\theta>-\alpha),  
\end{equation} 
which is well suited to describe diffusion on fractals \cite{osh}. It is encountered 
in geology where a power-law distribution of fracture 
lengths is responsible for transport in a rock \cite{pai}; this self-similar structure of 
a fracture and fault network is characterised by a fractal dimension and determines a pattern of water 
and steam penetration in the rock \cite{taf}. The parameter $\theta$ estimates a degree of the memory 
nonhomogeneity: for $\theta<0$, $g(x)$ is small near the origin which means long waiting times whereas 
for $\theta>0$ transport is inhibited at large distances. 
The deterministic force is linear when $\bar\gamma=1+\theta$ and Eq.(\ref{frace}) takes the form, 
\begin{equation}
\label{fraceo}
\frac{\partial p_0(x,\tau)}{\partial \tau}=\lambda\frac{\partial}{\partial x}[x p(x,\tau)]+
\frac{\partial^\alpha[|x|^{-\theta}p_0(x,\tau)]}{\partial|x|^\alpha}, 
\end{equation}
that leads, after the Fourier transformation, to the equation 
\begin{equation}
\label{fraceko}
\frac{\partial}{\partial \tau}{\widetilde p_0}(k,\tau)=-\lambda k\frac{\partial}{\partial k}
{\widetilde p_0}(k,\tau)-|k|^\alpha{\cal F}[|x|^{-\theta}p_0(x,\tau)].
\end{equation} 
Eq.(\ref{fraceo}) cannot be exactly solved for arbitrary $x$; we look for a solution in the limit 
of small $|k|$. It can be represented by a stable distribution but only in a limited range 
of $\theta$ \cite{sro14a}. The solution valid for an arbitrary large $\theta$ contains a power-law 
tail but its shape at small $|x|$ is different from that for the stable distribution \cite{sro09a}. 
In this paper, we assume that the solution has a scaling form 
that, in the asymptotic regime, is defined by the parameter $\alpha$, 
\begin{equation}
\label{solo}
p_0(x,\tau)=a(\tau)p_0(a(\tau)x)=a(\tau)^{-\alpha}|x|^{-\alpha-1}, 
\end{equation}
and the validity of a tentative assumption (\ref{solo}) will be verified 
and the unknown function $a(\tau)$ determined by inserting Eq.(\ref{solo}) to Eq.(\ref{fraceko}). 
For that purpose, both Fourier transforms in Eq.(\ref{fraceko}) 
have to be evaluated in a limit of small $|k|$. Transform in the last term is given by the integral (\ref{ost}), 
\begin{equation}
\label{osto}
a(\tau)\int|x|^{-\theta}p_0(a(\tau)x)dx=a(\tau)^\theta\int|x|^{-\theta}p_0(x)dx\equiv a(\tau)^\theta h_0,
\end{equation}
where $h_0=$const. By the evaluation of the Fourier transform from $p_0(x,\tau)$ we apply 
the same procedure as in Sec.II and obtain 
\begin{equation}
\label{ftr1o}
{\widetilde p_0}(k,\tau)=1-\frac{a(\tau)^{-\alpha}\pi}{\Gamma(\alpha+1)\sin(\alpha\pi/2)}|k|^{\alpha}. 
\end{equation}
Inserting the above expression to Eq.(\ref{fraceko}) produces a differential equation, 
\begin{equation}
\label{rroz}
\dot a(\tau)-\lambda a+\frac{h_0}{\alpha c}a(\tau)^{\alpha+\theta+1}=0,
\end{equation}
which can be solved by separation of the variables: 
\begin{equation}
\label{asol}
a(\tau)=A\left[1-\hbox{e}^{-\lambda(\alpha+\theta)\tau}\right]^{-1/(\alpha+\theta)}; 
\end{equation}
in this equation, $A=(\alpha\lambda c/h_0)^{1/(\alpha+\theta)}$ and $c$ is defined by Eq.(\ref{gor}). 
The final asymptotic solution is given 
by Eq.(\ref{solo}) and this expression is exact in respect to time- and position-dependence; 
the constant $h_0$ contains details of the solution for all $x$ and cannot be determined by the above 
procedure. The case $\theta=0$ is exceptional: then Eq.(\ref{fraceo}) can be solved exactly 
for arbitrary $x$ and the solution resolves itself to the stable distribution \cite{jes}.

\section{Relaxation to stationary states}

Stochastic properties of the systems for which particle is subjected to the Gaussian white noise 
and remains under the influence of an external deterministic force are usually described by 
the Fokker-Planck equation. The expansion of its solution to the eigenfunctions of 
the Fokker-Planck operator contains single modes which decay exponentially in time 
to a stationary state representing a Gibbs-Boltzmann 
distribution and the temperature is related to the diffusion coefficient by the fluctuation-dissipation 
theorem \cite{met}. Generalisation of that equation takes into account the memory effects by introducing 
a fractional derivative over time (the fractional Fokker-Planck equation). In this generalised description, 
a different relaxation pattern of the single modes emerges: the decay is governed by the Mittag-Leffler function 
\cite{met} and, in the case of the harmonically bound particle, also the variance decays 
according to that pattern \cite{met99}. The Mittag-Leffler relaxation naturally emerges 
in the subordination technique \cite{pir}. 

The system, relaxation properties of which are analysed in the present paper, is defined by the stochastic 
equation (\ref{la}). It includes the random force in a form of the L\'evy stable distribution 
and the non-Markovian nature of the process is taken into account by applying a subordination technique: 
the process -- described in Sec.II and given by $p_0(x,\tau)$ -- is subordinated to the random time. 
This time is defined by a one-sided, maximally asymmetric stable L\'evy distribution 
$h'(\tau,t)=L_\beta(\tau)$, where $0<\beta<1$, which vanishes for $\tau<0$. Since the processes 
$x(\tau)$ and $\tau(t)$ are statistically independent, the distribution 
as a function of the physical time is given by the total probability formula, 
\begin{equation}
\label{inte}
p(x,t)=\int_0^\infty p_0(x,\tau)h(\tau,t)d\tau, 
\end{equation} 
where $h(\tau,t)=\frac{t}{\beta\tau}L_\beta(\frac{t}{\tau^{1/\beta}})$ denotes an inverse subordinator \cite{pir}. 
In the next subsection, we derive that distribution for the linear case. Note that stationary state 
is completely determined by the distribution $p_0(x,\tau)$, i.e. by the noise type and the potential, 
whereas the subordinator, as a source of the subdiffusion, influences only the relaxation to that state. 

\subsection{Linear case} 

Let us consider the case $\bar\gamma=1+\theta$ which condition means that the effective deterministic 
force in the first equation (\ref{las}) is linear and the subordinated process is governed by Eq.(\ref{fraceo}). 
Note that then the force $F_d$ in Eq.(\ref{dfor}) may be nonlinear. 
The density $p(x,t)$ follows from Eq.(\ref{inte}) and can be evaluated by a direct integration. 
For that purpose, we insert the expression (\ref{ftr1o}) to Eq.(\ref{inte}) and calculate the Fourier 
and Laplace transform taking into account that the Laplace transform from $h(\tau,t)$ is given by 
$\bar h(\tau,u)=u^{\beta-1}\exp(-\tau u^\beta)$. This procedure yields 
\begin{equation}
\label{p1}
{\widetilde p}(k,u)=\frac{1}{u}-c|k|^\alpha u^{\beta-1}\int_0^\infty a(\tau)^{-\alpha}\hbox{e}^{-\tau u^\beta}d\tau 
\end{equation}
and the expression for $a(\tau)^{-\alpha}$ results from Eq.(\ref{asol}) by applying a direct exponentiation: 
\begin{equation}
\label{adoal}
A^\alpha a(\tau)^{-\alpha}=1+\sum_{j=1}^\infty\frac{(-1)^j}{j!}\hbox{e}^{-\lambda(\alpha+\theta)j\tau}
\prod_{i=0}^{j-1}(q-i), 
\end{equation}
where $q=\alpha/(\alpha+\theta)$. The integration yields the transformed density, 
\begin{equation}
\label{p2}
{\widetilde p}(k,u)=\frac{1}{u}-cA^{-\alpha}|k|^\alpha\left[\frac{1}{u}+
\sum_{j=1}^\infty\frac{(-1)^j}{j!}\frac{u^{\beta-1}}{u^\beta+\lambda(\alpha+\theta)j}\prod_{i=0}^{j-1}(q-i)\right].
\end{equation}
The above transform can easily be inverted when we realise that the $u-$dependent term corresponds to 
a Mittag-Leffler function that is defined by the series 
\begin{equation}
\label{mlf}
E_\beta(x)=\sum_{n=0}^\infty\frac{x^n}{\Gamma(1+\beta n)} 
\end{equation}
and resolves itself to an ordinary exponential for $\beta=1$. The final expression for the 
asymptotic form of the density in the Fourier space reads, 
\begin{equation}
\label{pf}
{\widetilde p}(k,t)=1-cA^{-\alpha}[1-\chi(t)]|k|^\alpha, 
\end{equation}
where a function 
\begin{equation}
\label{chiodt}
\chi(t)=-\sum_{j=1}^\infty\frac{(-1)^j}{j!}E_\beta[-\lambda(\alpha+\theta)jt^\beta]\prod_{i=0}^{j-1}(q-i),
\end{equation}
involving a series of the Mittag-Leffler functions, determines the relaxation to the stationary state and $\chi(\infty)=0$. 
Inversion of the Fourier transform yields a time-dependence of the asymptotic solution, 
\begin{equation}
\label{pas}
p(x,t)=A^{-\alpha}[1-\chi(t)]|x|^{-\alpha-1}~~~~(|x|\gg1). 
\end{equation}
Notice that $\chi(t)$ does not depend on $h_0$ and is exact, 
in contrast to the intensity of $p(x,t)$; the latter quantity depends on the shape of the density at small $|x|$ 
and is not determined by the present method. 
The long-time behaviour of the relaxation function $\chi(t)$ agrees with 
that of the Mittag-Leffler function and then $\chi(t)\propto t^{-\beta}$ but the effective relaxation time 
may strongly depend on the parameter $\theta$. Fig.1 presents $\chi(t)$ calculated from Eq.(\ref{chiodt}) 
for a few values of $\theta$. The Mittag-Leffler function was determined from Eq.(\ref{mlf}) for small $t$ and 
from the expansion 
\begin{equation}
\label{mlfd}
E_\beta(x)=-\sum_{n=1}^\infty\frac{x^{-n}}{\Gamma(1-\beta n)} 
\end{equation}
for large $t$; the convergence could not be achieved for intermediate values \cite{uwa}. Despite the same asymptotic 
form, the relaxation time substantially differs for various $\theta$: it is very large for $\theta<0$. 
\begin{center}
\begin{figure}
\includegraphics[width=80mm]{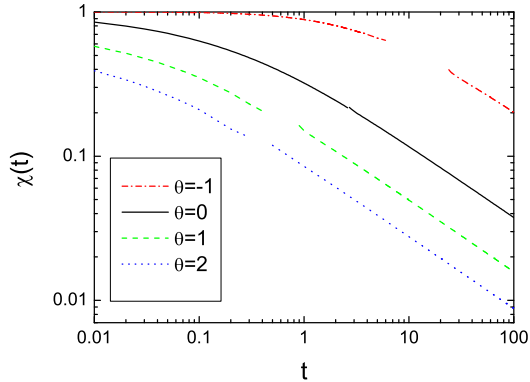}
\caption{(Colour online) Relaxation function $\chi(t)$ calculated from Eq.(\ref{chiodt}) for $\alpha=1.5$, 
$\beta=0.5$, $\lambda=1$ and a few values of $\theta$.}
\end{figure}
\end{center} 

The time-dependent density distribution for a stochastic dynamical system satisfies a Fokker-Planck 
equation which, in the presence of jumps and/or long waiting times, is fractional. This equation 
can be derived from the generalised master equation and CTRW \cite{met}; it has been applied, 
in particular, to describe chaotic systems characterised by 
a hierarchical structure of the trajectories \cite{zas}. The equation in the latter case possesses, 
in general, a variable diffusion coefficient. The fractional Fokker-Planck equation has a stochastic 
representation in a form of equations similar to (\ref{las}). This has been proved for 
the homogeneous case with an arbitrary potential \cite{mag} and that equivalence can be 
generalised to the nonhomogeneous case. Indeed, we demonstrate in the Appendix A 
that the function $p(x,t)$, Eq.(\ref{pas}), satisfies the following Fokker-Planck equation, 
\begin{equation}
\label{glef}
\frac{\partial p(x,t)}{\partial t}={_0}D_t^{1-\beta}
\left[\lambda\frac{\partial}{\partial x}[x p(x,t')]+
\frac{\partial^\alpha}{\partial|x|^\alpha}(|x|^{-\theta} p(x,t))\right],
\end{equation} 
where the Riemann-Liouville fractional derivative is defined by the integral operator, 
\begin{equation}
\label{rlo}
_0D_t^{1-\beta}f(t)=\frac{1}{\Gamma(\beta)}\frac{d}{dt}\int_0^t dt'\frac{f(t')}{(t-t')^{1-\beta}}.  
\end{equation} 

\subsection{Nonlinear case: a numerical analysis}

In this subsection, we consider the general case when the effective deterministic 
force in the first equation (\ref{las}) is a nonlinear function of $x$ ($\bar\gamma\ne 1+\theta$). 
To evaluate the density $p(x,t)$ from Eq.(\ref{inte}) we need the distribution for the subordinated 
process, $p_0(x,\tau)$, i.e. the solution of Eq.(\ref{frace}). This distribution 
cannot be found analytically, even for $\theta=0$, if $\gamma=\bar\gamma-\theta\ne1$. 
Therefore, we must resort to a numerical analysis details of which are presented in Appendix B. 
\begin{center}
\begin{figure}
\includegraphics[width=80mm]{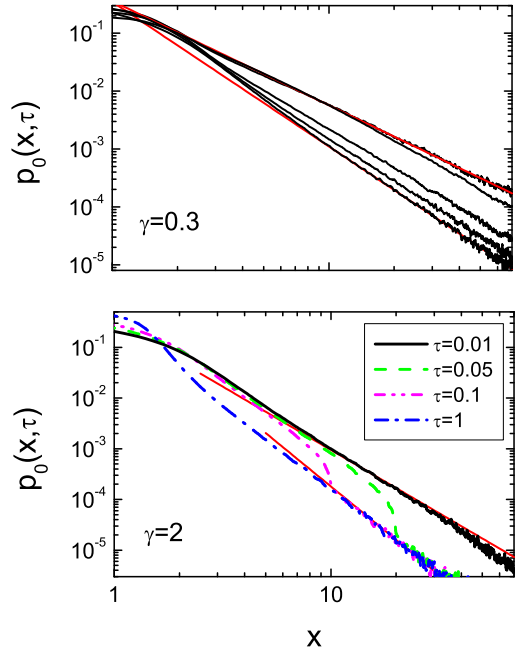}
\caption{(Colour on line) The time-evolution of the density distribution for the subordinated process corresponding to 
two values of $\gamma$ and calculated for $\alpha=1.5$ and $\theta=1$ from Eq.(\ref{las1}). The curves in the upper part 
correspond to $\tau=$0.05, 0.3, 1, 5 and 20 (from bottom to top). The straight red lines mark 
the dependence $x^{-1-\alpha}$ and $x^{-\alpha-\gamma}$. Averaging was performed over $10^7$ trajectories.}
\end{figure}
\end{center} 
First, we consider Eq.(\ref{las1}) to get a qualitative picture of the solutions for various values of 
$\gamma$. Just a simple analysis of Eq.(\ref{las1}) indicates that the cases $\gamma>1$ and $\gamma<1$ 
must be different. 
The particle, initially positioned at $x=0$, is subjected only to the random force at the beginning 
of the time-evolution since the first term in Eq.(\ref{las1}) 
can be neglected for $\gamma\ge0$. Hence the density distribution corresponds 
to the case without any potential and $p_0(x,\tau)\propto|x|^{-\alpha-1}$. However, 
the density approaches the stationary limit, 
which is of the form $|x|^{-\alpha-\gamma}$, differently for both cases: the tail becomes steeper ($\gamma>1$) 
or flatter ($\gamma<1$) during the evolution. The density distributions 
for those cases are presented, as a function of time, in Fig.2; we assume, from now on, $\lambda=1$. 
If $\gamma$ is large, the asymptotic distribution, $|x|^{-\alpha-\gamma}$, is visible already 
at a very short time but restricted to a large distance. The stationary distribution is reached at $\tau=1$ 
and the curves for the intermediate cases show a widening of the stationary tail. In contrast to this picture, 
the case $\gamma=0.3$ exhibits a gradual change of the slope for the entire tail and the stationary 
distribution is reached at $\tau=10$. 
\begin{center}
\begin{figure}
\includegraphics[width=80mm]{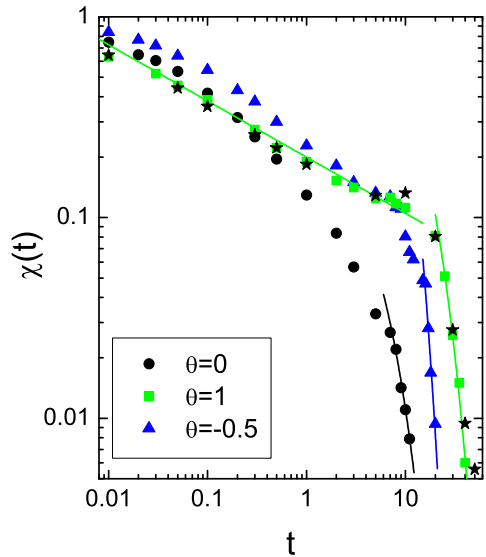}
\caption{(Colour on line) The relaxation function for $\alpha=1.9$, $\beta=0.5$, $\bar\gamma=3$ and 
three values of $\theta$. The solid lines on the right side mark an exponential, 
$\exp(-c_\chi t)$, with $c_\chi=0.33$, 0.4 and 0.14 (from left to right). The solid line 
on the left side, fitted to the data for $\theta=1$, marks a function $t^{-0.28}$. Stars 
correspond to the case $\bar\gamma=4$ and $\theta=1$. Averaging was performed over $10^4$ trajectories.}
\end{figure}
\end{center} 

We consider both cases separately and assume, for the large $\gamma$, the finite variance: 
$\alpha+\gamma>3$. Then 
\begin{equation}
\label{varch}
{\widetilde p}(k,t)\approx 1-a(t)k^2=1-a_e k^2(1-\chi(t)), 
\end{equation}
where $a(t)$ is a function describing a time-dependence of the asymptotic density which converges 
to its stationary value $a_e=a(\infty)$; on the other hand, we have $a(0)=0$ for the initial condition in the 
form of the delta function. The relaxation function 
\begin{equation}
\label{chinl}
\chi(t)=1-a(t)/a_e=1-\langle x^2\rangle(t)/\langle x^2\rangle(\infty)
\end{equation}
will serve as an estimator of the convergence rate to the stationary state. It is shown in Fig.3 
for a few values of $\theta$. In contrast to the linear case, the relaxation is not smooth: 
it exhibits two regions of a very different slope. At a short time, $\chi(t)$ diminishes relatively 
slowly whereas at large $t$ it rapidly falls according to the exponential rate. This picture emerges for both 
positive and negative $\theta$ but the break down of the curve is especially pronounced 
for $\theta=-0.5$. The case $\theta=0$ also indicates the exponential fall but the difference between 
the behaviour for small and large $t$ is less striking. The decline at small $t$ 
and for $\theta\ne0$ obeys, approximately, a power-law dependence; the slope is more gentle 
than for the linear case. We conclude that instead of the power-law pattern, present for the linear case, 
the exponential relaxation is observed at long times. The curves in Fig.3 correspond to different effective 
potentials in Eq.(\ref{las}) and one can ask whether this is a reason of their different shapes. To check that, 
we performed a calculation for $\bar\gamma=4$ and $\theta=1$ which corresponds to the same effective potential 
as the case $\bar\gamma=3$ and $\theta=0$. Results, presented in Fig.3, indicate a clear disagreement 
of those cases and the relaxation pattern is determined by $\theta$: the curves for the parameter sets 
$\bar\gamma=4$, $\theta=1$ and $\bar\gamma=3$, $\theta=1$ actually coincide. 
\begin{center}
\begin{figure}
\includegraphics[width=80mm]{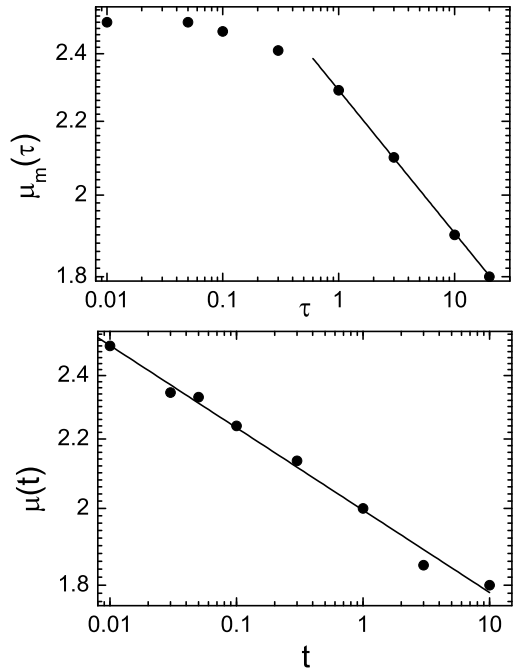}
\caption{Slope of the density $p_0(x,\tau)$ calculated from Eq.(\ref{las1}) for $\alpha=1.5$, $\beta=0.5$, 
$\gamma=0.3$ and $\theta=1$ (upper part). The lower part presents the slope as a function of the physical time, 
calculated from Eq.(\ref{inten}). Lines mark the functions $\tau^{-0.08}$ and $t^{-0.049}$ 
for the upper and lower part, respectively. Each point was obtained from the density evaluated by 
averaging over $10^7$ trajectories.}
\end{figure}
\end{center} 

Estimation of the decay rate to the stationary state in terms of the function $\chi(t)$, Eq.(\ref{chinl}), 
is not possible for $\gamma<1$ ($\bar\gamma<\theta+1$) since then the variance is infinite. 
Instead, we consider the time-evolution of the slope of the asymptotic distribution, $\mu(t)$. 
Evaluation of the tails is numerically difficult -- in respect to the 
precision and the calculation time -- if one directly solves 
Eq.(\ref{la}). On the other hand, the distributions in the operational time for the process 
defined by Eq.(\ref{las}) are easier to handle 
and the tails can be precisely determined. Therefore, we calculated $\mu(t)$ from Eq.(\ref{las}). 
As the first step, the slope $\mu_m(\tau)$ and the relative intensity of $p_0(x,\tau)$, $c_m(\tau)$, were evaluated 
from Eq.(\ref{las1}); the densities are presented in the upper part of Fig.2. Knowing those quantities, 
the final asymptotic distribution $p(x,t)$ can be obtained by a direct integration by means of Eq.(\ref{inte}), 
\begin{equation}
\label{inten}
p(x,t)=\int_0^\infty c_m(\tau)|x|^{-\mu_m(\tau)}h(\tau,t)d\tau, 
\end{equation}
and we restrict our analysis to the case $\beta=1/2$ for which the inverse subordinator has a simple form, 
\begin{equation}
\label{gau}
h(\tau,t)=\frac{1}{\sqrt{\pi t}}\exp(-\frac{\tau^2}{4t}). 
\end{equation}
Slopes of the asymptotic distributions characterising both processes are presented in Fig.4. 
They have different shapes as functions of time: $\mu_m(\tau)$ is almost constant at 
small times and declines as a power-law reaching its stationary value at $\tau=20$. The integration 
(\ref{inten}) brings about a smoothing effect: $\mu(t)$ has a power-law shape for all times and the index 
$\mu$ is smaller than for the subordinated process at a corresponding time; the stationary state 
is reached at $t=10$. 

\section{Summary and conclusions}

We have considered a dynamical process defined by the stochastic stimulation -- governed by 
the L\'evy stable statistics -- and the external potential in the form of both harmonic 
and strongly non-linear oscillator. 
It has been assumed that the particle may be trapped for a long time and the trapping time 
varies with the position, as a result of a nonhomogeneous medium structure. 
The problem is formulated in terms of the subordination technique where the 
intensity of the random time in the subordinator is position-dependent. 
The asymptotics of the stationary state has been derived for the case of the non-linear 
deterministic force. It has a power-law form; the slope depends on both the stability 
index of the driving noise and the potential but is independent of the memory characteristics: 
$\beta$ and $\theta$. The relaxation process to that stationary state has been analysed for both 
harmonic and non-linear case. For the linear case (the effective potential is quadratic),  
the decay rate $t^{-\beta}$ is observed at long times but 
the value of the relaxation function $\chi$ strongly varies as a function of $\theta$; long 
trapping times near the origin ($\theta<0$) correspond to a long time of approaching 
the stationary state. In contrast to the case $\theta=0$, where relaxation obey the Mittag-Leffler 
pattern, $\chi(t)$ for the general case involves a linear combination of the Mittag-Leffler functions. 
The density distribution for the linear case satisfies the integro-differential equation with a complicated 
kernel that resolves itself to a fractional equation with a variable diffusion coefficient in a limit 
of long time. 

The density distributions for a nonlinear effective force exhibit a complicated and differentiated structure 
which qualitatively depends on the potential slope. 
The strong potentials mean a rapid development of the stationary slope whereas the weak potentials are characterised 
by a gradual flattening of the entire tail. In the former case, the relaxation function 
is always exponential for long times: the Mittag-Leffler pattern breaks down and the Debye relaxation 
is observed. At smaller times, the decay is slow and weakly depends on $\theta$. Numerical calculations 
indicate that the relaxation pattern is sensitive rather on the spacial nonhomogeneity of the memory, 
parametrised by $\theta$, than on a slope of the effective deterministic force. 

The results indicate that the relaxation for a complex system -- in which long tails of the random 
force distribution coexist with a position-dependent trap distribution -- to a stationary state depends on 
both an external potential and a structure of the trap positions. When the non-exponential 
pattern is observed, as it is the case for the linear systems, the latter characteristics may 
strongly influence the relaxation time. However, the linearity property refers not to 
the external potential alone but results from an interplay between this potential and the trap 
distribution. Therefore, though the nonlinearity changes the relaxation pattern to the exponential, 
presence of the Mittag-Leffler pattern in the experimental data does not necessarily imply 
that the linear deterministic force would be correct in a dynamical description of the case 
of the uniform trap distribution. Finally, we emphasise the non-homogeneous trap distribution 
can make the variance of the process value finite. The divergent 
variance for the L\'evy flights is often unphysical and we demonstrated that this difficulty may not 
emerge even for relatively weak potentials if $\theta$ is negative and sufficiently small.

\section*{APPENDIX A}

\setcounter{equation}{0}
\renewcommand{\theequation}{A\arabic{equation}}  

In the Appendix A, we demonstrate that the density $p(x,t)$, Eq.(\ref{pas}), satisfies 
the Fokker-Planck equation (\ref{glef}). 

 Let us consider the following integro-differential equation, 
\begin{equation}
\label{A.1}
\frac{\partial p(x,t)}{\partial t}=\int_0^t K(t-t')\left[\lambda\frac{\partial}{\partial x}[x p(x,t')]+
\frac{\partial^\alpha}{\partial|x|^\alpha}(|x|^{-\theta} p(x,t'))\right]dt', 
\end{equation} 
and take the Fourier-Laplace transform which we rewrite as $F_1=\bar K(u)(F_2+F_3)$. 
Next, we apply Eq.(\ref{p2}) and obtain 
\begin{equation}
\label{A.2}
F_1={\cal L}[\frac{\partial}{\partial t}{\widetilde p}(k,t)]=-cA^{-\alpha}|k|^\alpha\left[1+
u^\beta\sum_{j=1}^\infty\frac{(-1)^j}{j!}\frac{\prod_{i=0}^{j-1}(q-i)}{u^\beta+\lambda(\alpha+\theta)j}\right]. 
\end{equation} 
Similarly, the first term on rhs reads 
\begin{equation}
\label{A.3}
F_2=-\lambda k\frac{\partial}{\partial k}{\widetilde p}(k,u)=
c\lambda\alpha A^{-\alpha}|k|^\alpha\left[\frac{1}{u}+
u^{\beta-1}\sum_{j=1}^\infty\frac{(-1)^j}{j!}\frac{\prod_{i=0}^{j-1}(q-i)}{u^\beta+\lambda(\alpha+\theta)j}\right]
\end{equation}
and the last one, 
\begin{equation}
\label{A.4}
F_3=-|k|^\alpha{\cal F}[|x|^{-\theta}\bar p(x,u)]=
-|k|^\alpha h_0 u^{\beta-1}\int_0^\infty a^\theta\hbox{e}^{-\tau u^\beta}d\tau=
-|k|^\alpha h_0 A^\theta \left[\frac{1}{u}+
u^{\beta-1}\sum_{j=1}^\infty\frac{(-1)^j}{j!}\frac{\prod_{i=0}^{j-1}(q'-i)}{u^\beta+\lambda(\alpha+\theta)j}\right], 
\end{equation}
where $q'=-\theta/(\alpha+\theta)=q-1$. Denoting the sum in Eq.(\ref{A.2}) by $S_1$,
\begin{equation}
\label{A.5}
S_1=\sum_{j=1}^\infty\frac{(-1)^j}{j!}\frac{\prod_{i=0}^{j-1}(q-i)}{u^\beta+\lambda(\alpha+\theta)j}
\end{equation}
and, similarly, the sum in Eq.(\ref{A.3}) by $S_2$, we get 
\begin{equation}
\label{A.6}
F_2+F_3=c\lambda\alpha A^{-\alpha}|k|^\alpha u^{\beta-1}(S_1-S_2);
\end{equation}
in the above equation we took into account that $A^\theta=\frac{c\lambda\alpha}{h_0}A^{-\alpha}$. 
The expression $S_1-S_2$ can be simplified by changing the product index in the second sum: 
\begin{equation}
\label{A.7}
S_1-S_2=\sum_{j=1}^\infty\frac{(-1)^j}{j!}\frac{1}{u^\beta+\lambda(\alpha+\theta)j}\prod_{i=0}^{j-1}(q-i)(1-\frac{q-j}{q})=
q^{-1}\sum_{j=1}^\infty\frac{(-1)^j}{(j-1)!}\frac{\prod_{i=0}^{j-1}(q-i)}{u^\beta+\lambda(\alpha+\theta)j}\equiv
q^{-1}S.
\end{equation}
Dividing $F_1$ by $F_2+F_3$ yields the expression for the Laplace transform from the kernel, 
\begin{equation}
\label{A.8}
\bar K(u)=-\frac{1}{\lambda(\alpha+\theta)}\frac{1+S_1u^\beta}{S}u^{1-\beta},
\end{equation}
which can be evaluated by using the identity, 
\begin{equation}
\label{A.9}
\sum_{j=1}^\infty\frac{(-1)^j}{j!}\prod_{i=0}^{j-1}(q-i)=-1. 
\end{equation}
Finally, $\bar K(u)=u^{1-\beta}$, inversion of this transform produces the operator (\ref{rlo}) and 
Eq.(\ref{A.1}) takes a form of the fractional equation (\ref{glef}). 

\section*{APPENDIX B}

\setcounter{equation}{0}
\renewcommand{\theequation}{B\arabic{equation}}  

In the Appendix B, we present numerical details of the numerical evaluation of the stochastic 
trajectories from Eq.(\ref{la}). 

The time-evolution of a stochastic system which is represented 
by the subordination equations can be numerically evaluated, namely the process value as a function 
of the physical time can be found, without an explicit inversion of the subordinator \cite{wer1,kle}. 
We solve a coupled set of equations (\ref{la}) the first of which is an ordinary Langevin 
equation and can be discretized, 
\begin{equation}
\label{B.1}
x(\tau_n)=x(\tau_{n-1})+F_d[x(\tau_{n-1})]\Delta\tau+\Delta\eta\Delta\tau^{1/\alpha}. 
\end{equation}
The second equation (\ref{la}), in turn, represents a strictly increasing stable motion (since $g(x)\ge0$), 
defined as 
\begin{equation}
\label{B.2}
\tau(t)=\hbox{inf}\{\tau: t(\tau)\ge t\},  
\end{equation}
and the discretized equation reads 
\begin{equation}
\label{B.3}
t(\tau_n)=t(\tau_{n-1})+g(x(\tau_{n-1}))\Delta\xi\Delta\tau^{1/\beta}. 
\end{equation}
Increments of the stable distributions, $\Delta\eta$ and $\Delta\xi$, are sampled by means of the standard 
procedure \cite{wer}. 
Operationally, $x(T)$ follows from the integration of both equations as long as $t\ge T$; finally, 
the definition (\ref{B.2}) is applied. In the calculations presented in Fig.2-4, the step $\Delta\tau=10^{-4}$ 
was used.

\end{document}